\documentclass[pra,aps,twocolumn,superscriptaddress]{revtex4-2}

\usepackage{natbib}
\usepackage[T1]{fontenc}
\usepackage[latin9]{inputenc}
\usepackage{babel}
\usepackage{float}
\usepackage{amsmath}
\usepackage{amssymb}
\usepackage{amsfonts}
\usepackage{xcolor}
\usepackage{bbm}
\usepackage[caption=false,font=normalsize,labelfont=sf,textfont=sf]{subfig} % caption=true centers the caption. Looks horrible.
\usepackage{graphicx,epstopdf}
\usepackage{url}
\usepackage{textcase}
\usepackage{bm}
\usepackage[unicode=true]{hyperref}
\usepackage{listings}
\usepackage{physics}
\hypersetup{colorlinks=true, linkcolor=blue, filecolor=magenta, urlcolor=cyan, citecolor=red}

\newcommand{\PT}{{\mathcal {PT}}}

\newcommand{\al}{\alpha}

\newcommand{\beqa}{\begin{eqnarray}}
\newcommand{\eeqa}{\end{eqnarray}}

\renewcommand{\Im}{{\rm Im}}

\begin{document}
\title{$\mathcal{PT}$-symmetry breaking in a Kitaev chain with one pair of gain-loss potentials}
\author{Kaustubh S. Agarwal}
%\email{ksagawra@iupui.edu}
\affiliation{Department of Physics, Indiana University-Purdue University Indianapolis, Indianapolis 46202, Indiana, USA}
\author{Yogesh N. Joglekar}
 \affiliation{Department of Physics, Indiana University-Purdue University Indianapolis, Indianapolis 46202, Indiana, USA}
\date{\today}

\begin{abstract}
Parity-time ($\mathcal{PT}$) symmetric systems are classical, gain-loss systems whose dynamics are governed by non-Hermitian Hamiltonians with exceptional-point (EP) degeneracies. The eigenvalues of a $\mathcal{PT}$-symmetric Hamiltonian change from real to complex conjugates at a critical value of gain-loss strength that is called the $\mathcal{PT}$ breaking threshold. Here, we obtain the $\mathcal{PT}$-threshold for a one-dimensional, finite Kitaev chain---a prototype for a p-wave superconductor--- in the presence of a single pair of gain and loss potentials as a function of the superconducting order parameter, on-site potential, and the distance between the gain and loss sites. In addition to a robust, non-local threshold, we find a rich phase diagram for the threshold that can be qualitatively understood in terms of the band-structure of the Hermitian Kitaev mo del. In particular, for an even chain with zero on-site potential, we find a re-entrant $\mathcal{PT}$-symmetric phase bounded by second-order EP contours. Our numerical results are supplemented by analytical calculations for small system sizes.
\end{abstract}
\maketitle

%------------------------------------- Section 1 ----------------------------------------%

\section{\label{sec:Intro}Introduction}
%Intro on PT symmetry ...

A complex extension to quantum mechanics, based on a special class of non-Hermitian Hamiltonians with purely real spectra, was discovered more than two decades ago~\cite{bender1998,bender2002,ali2002,ali2010}. These continuum Hamiltonians on an infinite line shared the property that each of them is invariant under combined operations of parity and time-reversal ($\PT$); i.e. the Hamiltonian commutes with the $\PT$ operator. Over the past decade, it has become clear that $\PT$-symmetric systems represent open, classical systems with balanced, spatially or temporally separated gain and loss that are represented by complex real-space potentials~\cite{feng2017,el2018,ozdemir2019}.

The spectrum $E_\alpha$ of a $\PT$-symmetric Hamiltonian $H(\gamma)$ is real at small non-Hermiticities $\gamma$ and turns into complex conjugate pairs at large non-Hermiticities. When the spectrum is purely real, we can choose an eigenvector $\ket{\epsilon_\alpha}$ of $H$ to be a simultaneous eigenvector of the antilinear $\PT$-operator with eigenvalue +1. When the spectrum is complex, the $\PT$ operator acting on $\ket{E_\alpha}$ transforms it into the eigenvector with complex-conjugate eigenvalue, i.e. $\PT|E_\alpha\rangle=|E^*_\alpha\rangle$ where $*$ denotes complex conjugation. The transition from a purely real to complex conjugate spectrum occurs when $\gamma=\gamma_\mathrm{PT}$ where $\gamma_\mathrm{PT}$ is called the $\PT$-symmetry breaking threshold. At the threshold, the geometric multiplicity of the eigenvalues of the Hamiltonian $H(\gamma_\mathrm{PT})$ is smaller than the algebraic multiplicity. Such a Hamiltonian degeneracy is called an exceptional point (EP) degeneracy, where not only do the eigenvalues become degenerate, but the corresponding eigenvectors also coalesce. 

In the past decade $\PT$-symmetric systems with balanced gain-loss have been realized in classical wave systems including evanescently coupled waveguides~\cite{ruter2010}, fiber loops~\cite{regensburger2012}, optical resonators ~\cite{chang2014, hodaei2014}, electrical circuits ~\cite{schindler2011,choi2018}, and mechanical oscillators~\cite{bender2013}. However, since the EP degeneracies also occur for Hamiltonians with mode selective dissipation, the dynamics of $\mathcal{PT}$-symmetric Hamiltonians have also been realized in purely lossy classical systems consisting of coupled waveguides, resonators, or electrical circuits~\cite{guo2009,lenmontiel2018}, semiclassical systems with ultracold atoms~\cite{li2019}, and quantum systems~\cite{Wu2019,naghiloo2019,klauck2019}.

In this paper we obtain the $\PT$-breaking threshold in a one-dimensional Kitaev model in the presence of one pair of gain-loss potentials $\pm i\gamma$ located on reflection-symmetric sites. Kitaev model is a toy model for a topological superconductor with Majorana fermions as excitations. A Majorana fermion is, by construction, its own antiparticle, i.e. it is a fermion constructed from a electron-hole pair. There are a number of studies on topological superconductors with non-Hermitian, $\PT$-symmetric potentials that preserve the translational invariance of the system~\cite{wang2015,li2018,yao2018,yuce2016,klett2017,kawabata2018,li2020,sarma2015,leumer2020}. They focused on the fate of the edge states that are characterized by zero energy. A finite Kitaev chain with a pair of $\PT$-symmetric potentials at the edge has also been studied~\cite{yuce2016,kawabata2018,li2020,klett2017}. The result is the emergence of an additional pair of edge state with a non-zero energy eigenvalue~\cite{kawabata2018}. In contrast to these studies on the Majorana edge modes, here we concentrate on the variation of the $\PT$-breaking threshold $\gamma_\mathrm{PT}(m_0,N)$ with the location $m_0$ of the gain potential in a Kitaev chain of size $N$. 

This paper is organized as follows. In Sec.~\ref{sec:model} we describe the Hermitian Kitaev chain along with its symmetry properties, and then introduce the non-Hermitian perturbation. In Sec.~\ref{sec:results}, we present numerical results for the $\PT$-symmetry breaking threshold as a function of different parameters of the Hermitian model and the relative location of the gain potential. We point out key differences among the $\PT$-symmetry breaking thresholds for various settings of the on-site potentials $\mu$ and superconducting coupling strengths $\delta$. In Sec.~\ref{sec:N5case} we describe a small-system case of $N=5$ sites and analytically obtain the dependence of the $\PT$ threshold when the gain-loss potentials are farthest apart and closest together. In Sec.~\ref{sec:reentrant}, we show that the Kitaev model shows re-entrant $\mathcal{PT}$-symmetric phase, and map out its EP contours. Finally in Sec.~\ref{sec:conclusion}, we conclude by summarizing the results. 

%----------------------------------------- Section 2 ------------------------------------%

\section{\label{sec:model}Tight-binding Model}

The Kitaev model of a one dimensional, p-wave superconducting chain with $N$ sites and open boundary conditions is described by the following Hermitian Hamiltonian, 
\begin{eqnarray}
\label{eq:H_fermion}
    H_0 & = & - \mu \sum_{n=1}^N c_n^{\dagger}c_n -J \sum_{n=1}^{N-1} (c_{n}^{\dagger}c_{n+1} + \mathrm{h.c.})\nonumber\\
&+  &i\delta \sum_{n=1}^{N-1}(c_{n}c_{n+1} - \mathrm{h.c.}).
\end{eqnarray}
Here $c^{\dagger}_n$ and $c_n$ are fermionic creation and annihilation operators for site $n$ in the chain, $\mu$ is the on-site potential, $J>0$ is the nearest-neighbor hopping strength, and $\delta>0$ is the amplitude of the (p-wave) superconducting coupling for a Cooper pair that is localized across neighboring sites~\cite{kitaev2001}. The global phase of the superconducting order parameter is fixed at $\pi/2$ to ensure that Eq.(\ref{eq:H_fermion}) is parity-time symmetric, with the parity operator given by 
$\mathcal{P}: c_n\rightarrow c_{\bar{n}}$ where $\bar{n}=N+1-n$ is the mirror-symmetric counterpart of site $n$ and the time-reversal operator is given by complex conjugation, $\mathcal{T}=*$. 

We rewrite Eq.(\ref{eq:H_fermion}) by using the Bogoliubov-de Gennes representation in terms of the operator-vector $\Psi=(c_1,c^{\dagger}_1,c_2,c^{\dagger}_2 \dots c_N, c^{\dagger}_N)^T$ as 
$H_0=\Psi^\dagger H_\mathrm{BdG}\Psi$ where the $2N\times 2N$ matrix $H_\mathrm{BdG}$ in the site-representation is given by 
\begin{eqnarray}
H_\mathrm{BdG} & = & -\frac{\mu}{2} \sum_{n=1}^N\ket{n}\bra{n}\otimes \sigma_z \nonumber \\
        &   & -\frac{J}{2} \sum_{n=1}^{N-1}\left(\ket{n}\bra{n+1} + \ket{n+1}\bra{n}\right)\otimes \sigma_z \nonumber \\
        &   & + \frac{i\delta}{2} \sum_{n=1}^{N-1}\left(\ket{n}\bra{n+1} - \ket{n+1}\bra{n}\right)\otimes \sigma_x,
\label{Eq:kiHamil}
\end{eqnarray}
where $\sigma_x,\sigma_z$ are the standard Pauli matrices. For a chain with periodic boundary conditions, translational invariance allows us to transform the site-space Hamiltonian (\ref{Eq:kiHamil}) into momentum space, $\tilde{H}_\mathrm{BdG} =UH_\mathrm{BdG}U^{\dagger}$ 
with a unitary 
\begin{align}
U& =\frac{1}{\sqrt{N}} \sum_{k,n=1}^N e^{-i\ket{p_k}\bra{n}}\otimes\left( e^{i\pi/4}\mathbbm{1}_2+e^{-i\pi/4}\sigma_z\right).
\end{align}
The block-diagonalized momentum-space Hamiltonian is given by $\tilde{H}_\mathrm{BdG}=\sum_{k=1}^N h(p_k)|p_k\rangle\langle p_k|$ where $p_k=2\pi k/N$ are the discrete quasimomenta for a finite chain and 
\begin{eqnarray}
h(p) &=&
\begin{pmatrix}
-J\cos{p}-\mu/2 & -i\delta \sin{p} \\
i\delta \sin{p} & J\cos{p}+\mu/2
\end{pmatrix}.
\end{eqnarray}
The bulk energy spectrum of the Hamiltonian $\tilde{H}_\mathrm{BdG}$ is given by 
\begin{equation}
E_{\pm}(p)=\pm \sqrt{(J\cos{p} + \mu/2)^2 + \delta^2 \sin^2{p}},
\label{eq:bulk}
\end{equation}
and it shows that in the limit of an infinite chain, $N\gg 1$, the gap in the spectrum vanishes at $p=\pi$ when $\mu=2J$. For the finite chain, the spectrum Eq.(\ref{eq:bulk}) is symmetric about $\delta=0$ because $H_\mathrm{BdG}(-\delta)=\mathcal{S} H_\mathrm{BdG}(\delta)\mathcal{S}^\dagger$ with a unitary operator $\mathcal{S}=\mathbbm{1}_N\otimes\sigma_z$. When the boundary conditions are changed from periodic to open, in the zero chemical potential limit, Majorana zero-modes (fermionic excitations) appear localized on the edges of the chain. These edge modes are robust when gain-loss potentials are introduced on random sites~\cite{yuce2016} or on parity symmetric sites with disorder~\cite{kawabata2018}.

To this toy model with open boundary conditions, we add a pair of balanced gain-loss potentials 
$\pm i\gamma$ at mirror symmetric sites $m_0$ and $\bar{m}_0$,
\begin{equation}
i\Gamma= \frac{i\gamma}{2} \left(\ket{m_0}\bra{m_0} - \ket{\overline{m}_0}\bra{\overline{m}_0}\right)\otimes \sigma_z,
\end{equation}
and thereby get a non-Hermitian, $\mathcal{PT}$-symmetric Kitaev chain Hamiltonian 
\begin{equation}
\label{eq:hkitaev}
H_\mathrm{K}(\gamma,\delta,\mu) =H_\mathrm{BdG}+i\Gamma.
\end{equation}

% Figure 1--------------------------%
\begin{figure}[h]
\centering
\includegraphics[width=\columnwidth]{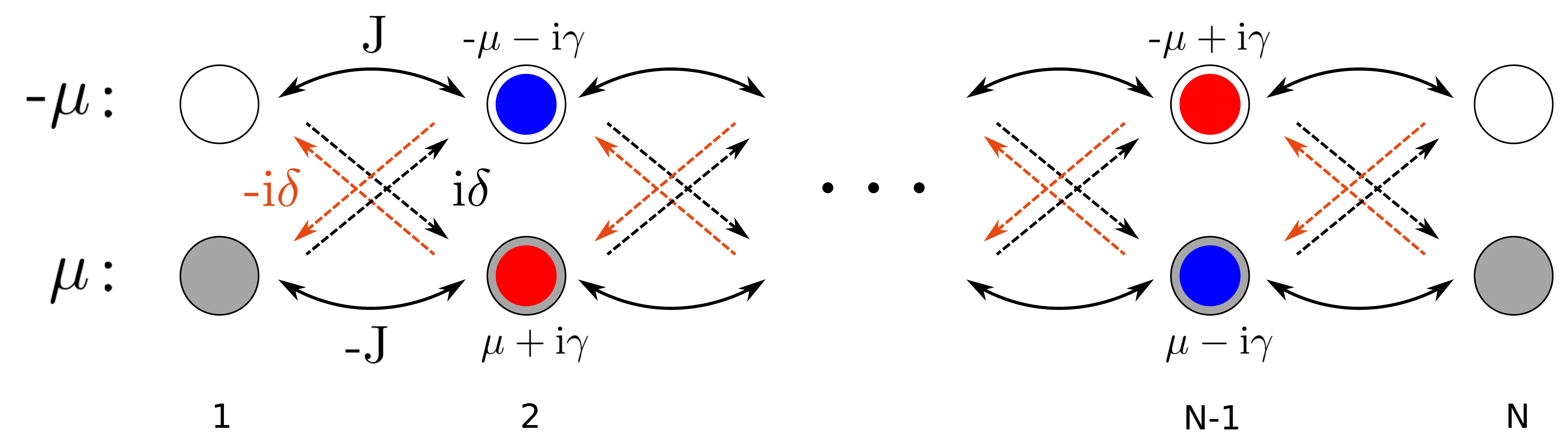}
\caption{Schematic representation of a Kitaev model with one pair of gain and loss potentials, Eq.(\ref{eq:hkitaev}). Two chains (gray and white) with detuned potentials $\pm\mu$ have nearest neighbor Hermitian tunneling amplitudes $\pm J$, and next-nearest-neighbor Hermitian amplitudes $\pm i\delta$. Due to the presence of two degree of freedom on each site~\cite{harsha2013}, the potential $i\gamma$ on site $m_0$ acts as gain (red) for one and loss (blue) for the other. This schematic can be realized with coupled resonator rings where one can engineer complex, Hermitian tunneling amplitudes.}
\label{fig:kichain}
\end{figure}
%----------------------------------%

Fig.~\ref{fig:kichain} shows schematic representation of a lattice model described by Eq.(\ref{eq:hkitaev}). Although the original model refers to many-body fermionic system with two bands, in its ``single-particle'' form, Eq.(\ref{eq:hkitaev}), it can be interpreted as two, detuned coupled chains with Hermitian nearest-neighbor couplings $\pm J$ and Hermitian, purely imaginary, next-nearest-neighbor couplings 
$\pm i\delta$~\cite{harsha2013}. The gain potential on site $m_0$, given by $(i\gamma/2) |m_0\rangle\langle m_0|\otimes\sigma_z$, then stands for gain in one chain and loss in the second chain. This representation of the $\mathcal{PT}$-symmetric Kitaev model can be experimentally implemented in resonator arrays where real and purely imaginary tunneling amplitudes can be easily engineered. In the next section we explore the global phase diagram for the $\mathcal{PT}$-symmetry breaking threshold $\gamma_\mathrm{th}(m_0,\delta,\mu)$. 

%----------------------------------- Section 3 ------------------------------------------%

\section{\label{sec:results} Numerical results for $\gamma_\mathrm{th}(m_0,\delta,\mu)$}

% Figure 2---------------------------%
\begin{figure*}
    \centering
    \includegraphics[width=0.9\textwidth]{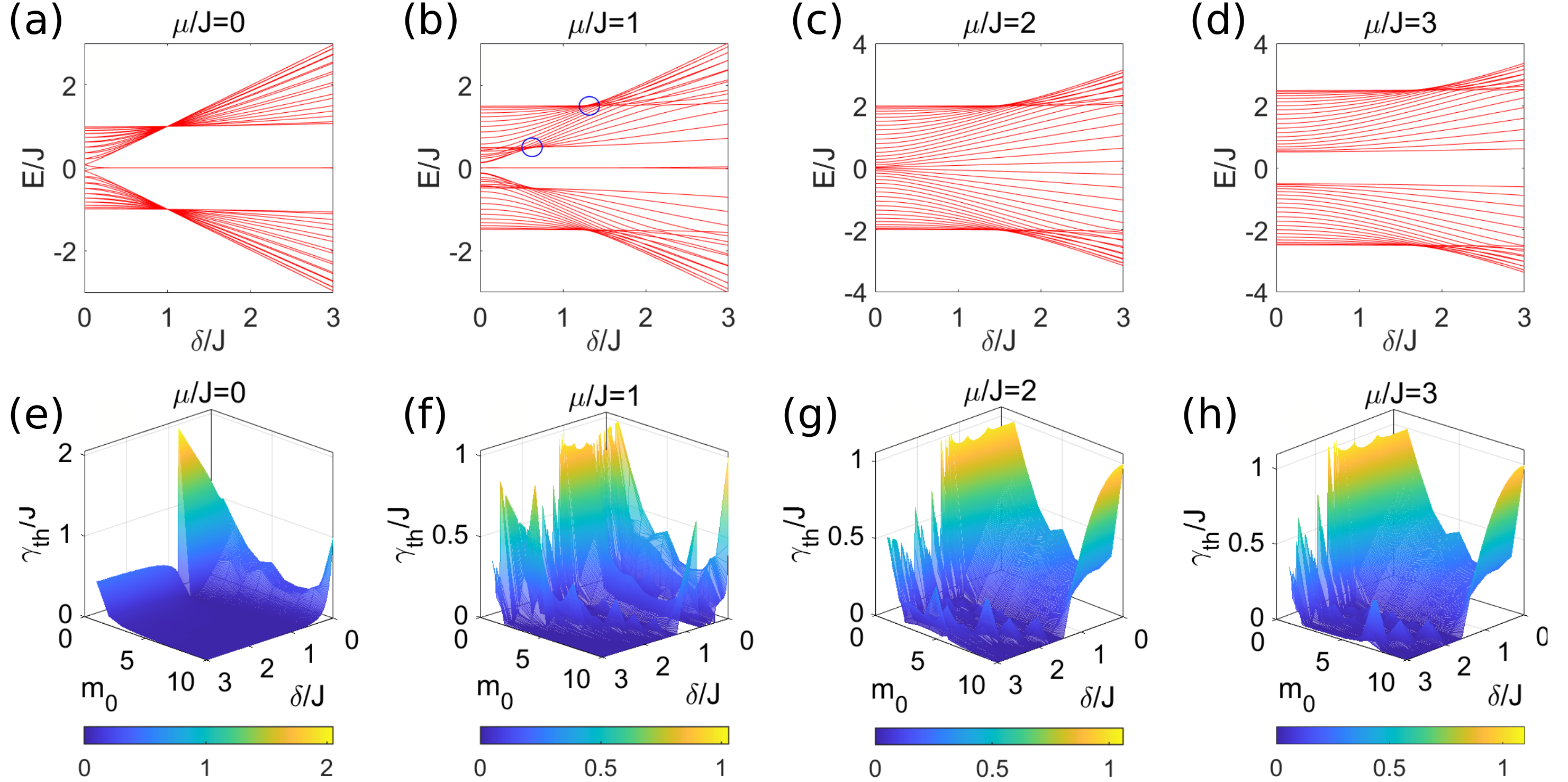}
    \caption{Energy eigenvalues (in units of $J$) of an $N=20$ site Hamiltonian $H_\mathrm{BdG}(\delta)$, Eq.(\ref{Eq:kiHamil}) as a function of detuning. (a) $\mu/J=0$, (b) $\mu/J=1$ have topological edge modes, while the system is in the topologically trivial phase at (c) $\mu/J=2$ and (d) $\mu/J=3$. Corresponding $\mathcal{PT}$ threshold values $\gamma_\mathrm{th}/J$ obtained from the Hamiltonian $H_\mathrm{K}$, Eq.(\ref{eq:hkitaev}) are plotted as a function of the gain location $m_0\in[1,N/2]$ and the superconducting order parameter $\delta/J$: (e) $\mu/J = 0$, (f)  $\mu/J = 1$, (g)  $\mu/J = 2$ (h)  $\mu/J = 3$. Most of these features can be understood in terms of Hermitian band structure, panels a-d.}
    \label{fig:PTeven}
\end{figure*}
%-----------------------------------%

% Figure 3----------------------------%
\begin{figure*}
    \centering
    \includegraphics[width=0.9\textwidth]{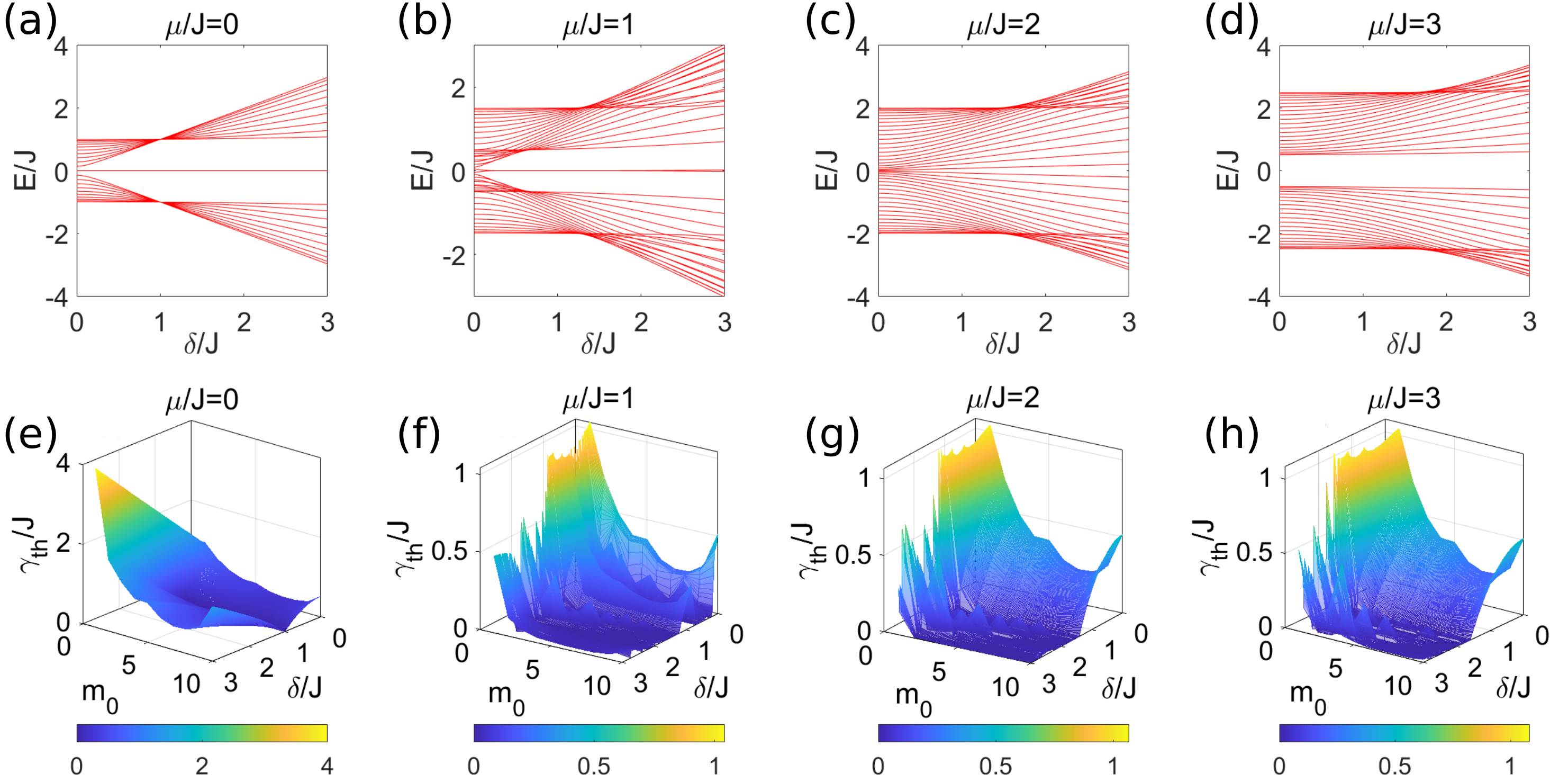}
    \caption{Energy eigenvalues (in units of $J$) of an $N=21$ site Hamiltonian $H_\mathrm{BdG}(\delta)$, Eq.(\ref{Eq:kiHamil}) as a function of detuning. (a) $\mu/J=0$, (b) $\mu/J=1$ have mid-gap states with zero energy, but these are not topological. (c) $\mu/J=2$ and (d) $\mu/J=3$ show emergence of a gapped spectrum. Corresponding $\mathcal{PT}$ threshold values $\gamma_\mathrm{th}/J$ obtained from the Hamiltonian $H_\mathrm{K}$, Eq.(\ref{eq:hkitaev}) are plotted as a function of the gain location $m_0\in[1,(N-1)/2]$ and the superconducting order parameter $\delta/J$: (e) $\mu/J = 0$, (f)  $\mu/J = 1$, (g)  $\mu/J = 2$ (h)  $\mu/J = 3$. Due to the absence of topological edge modes, the $\mathcal{PT}$ threshold behavior at $\mu=0$ is markedly different from that of an even chain, Fig.~\ref{fig:PTeven}(e). At nonzero detuning, the threshold is non-monotonically suppressed with increasing $\delta$.}
    \label{fig:PTodd}
\end{figure*}
%------------------------------------%

The results presented in this section are obtained by diagonalizing $H_\mathrm{K}$ or $H_\mathrm{BdG}$ for Kitaev chains of size $N=20$ (Fig.~\ref{fig:PTeven}) and $N=21$ (Fig.~\ref{fig:PTodd}). They remain qualitatively same for larger chain sizes, and the differences between even and odd parity chains persist in the large $N$ limit, as they do for a simple tight-binding model. All energy scales are measured in units of the tight-binding coupling $J=1$. Figures~\ref{fig:PTeven}(a)-(d) show the the energy eigenvalues $E_n$ for a Hermitian Kitaev chain as a function of the superconducting order parameter $\delta/J$. When $\delta=0=\mu$, we get the cosine-band of a tight-binding model. As the detuning $\delta$ is increased from (a) to (d), the two bands become well-separated. On the other hand, at a fixed detuning, when $\delta$ is increased, the bands develop fan-like linear dispersion, leading to massively degenerate flat bands at $\delta/J=1$ at zero detuning. As the detuning is increased from $\mu=0$, the system develops two crossing points (shown by blue circles in (b)). We also note that zero-energy states are present when $\mu<2J$. At $\mu/J=2$, the superconducting gap closes marking a phase transition to the topologically trivial phase. Here the mid-gap states become a part of the bulk. When $\mu$ is increased further, Fig.~\ref{fig:PTeven}(d), the system is in the trivial superconducting phase and energy spectrum is gapped. The topological, edge-localized zero energy states only occur when $\mu\leq 2J$. At $\mu=0$, these zero-energy states are fully localized on the end sites. When $\mu$ is increased, these states extend into the bulk of the chain, with an exponentially decaying probability density~\cite{leumer2020,sarma2015}. The Hermitian, (near or exact) degeneracies of the Hamiltonian $H_\mathrm{BdG}$ play an important role in determining the threshold gain-loss strength $\gamma_\mathrm{th}$ when a pair of gain-loss potentials is introduced at mirror symmetric sites. 

Figures~\ref{fig:PTeven}(e)-(h) show the numerically determined $\mathcal{PT}$-symmetry breaking threshold $\gamma_\mathrm{th}$ for the $N=20$ chain as a function of $m_0$ and superconducting order parameter $\delta$. When $\mu/J=0$ (panel e), we see that $\gamma_\mathrm{th}(m_0,\delta)$ has the characteristic U-shaped behavior~\cite{joglekar2010,agarwal2018} when $\delta=0$ and becomes mostly zero for intermediate locations $m_0\sim N/4$. When $m_0=1$, i.e. when the gain-loss locations are farthest apart, the $\mathcal{PT}$ threshold is maximized to $\gamma_\mathrm{th}=J$, and reflects the non-local robustness that is ubiquitous for systems with open boundary conditions~\cite{joglekar2010,Joglekar2013}. In this case, the states that participate in the $\mathcal{PT}$-breaking process are the mid-band states. As $m_0$ is increased, the threshold decreases and it rises back to $\gamma_\mathrm{th}=J$ when the gain and loss locations are nearest neighbors, i.e $m_0=N/2$. In this situation, all eigenvalues simultaneously and pairwise become complex, giving rise to maximal $\mathcal{PT}$-symmetry breaking~\cite{joglekar2011}. 

In contrast to the variation with $m_0$, we find that when the superconducting order parameter $\delta$ is varied, for most locations $m_0$, the $\mathcal{PT}$-threshold is uniformly suppressed from its $\delta=0$ value. The exception is the region $m_0\sim 1$, where, as $\delta$ is increased, we see that the $\mathcal{PT}$ threshold at $\delta=J$ is double its $\delta=0$ value~\cite{kawabata2018}, where the flat bands occur; see Fig.~\ref{fig:PTeven}(a). As $\delta$ is increased further, the threshold dips to zero and then increases reaching a steady, $\delta$-independent value of $\gamma_\mathrm{th}=J/2$. As the detuning $\mu$ is increased from zero, Fig.~\ref{fig:PTeven}(f), there is an overall suppression of the $\mathcal{PT}$-breaking threshold $\gamma_\mathrm{th}$ although the characteristic U-shape behavior as a function of $m_0$ and the non-monotonic behavior as a function of $\delta$ for farthest gain-loss potentials are both retained. These qualitative trends continue for $\mu\leq 2J$, i.e. when the system is in the topological phase. 

When the detuning is large, $\mu>2J$, the system enters trivial superconducting phase with no edge localized states, (g) and (h). In this regime, the system consists of two separated bands, and therefore the $\mathcal{PT}$-threshold does not sensitively depend on the detuning. On the other hand, when gain and loss are on nearest neighbor sites, $m_0=N/2$, the threshold is suppressed to zero for $\delta\sim 2J$. This is explained by the level crossings that occur near band edges; see Figs.~\ref{fig:PTeven}(c) and (d).

So are there any differences in the threshold behavior for an odd Kitaev chain? Figure~\ref{fig:PTodd} shows corresponding, representative results for a chain with $N=21$ sites. Panels (a)-(d) show the dispersions of the Hermitian Kitaev chain as a function of $\delta/J$ for increasing detuning values. At zero detuning, panel (a), the band structure looks similar to that in Fig.~\ref{fig:PTeven}(a), but with a key difference: there is no degenerate pair of topological zero energy states. As $\mu$ is increased, the qualitative evolution of the band structure is similar to that of an even Kitaev chain, with the band gap closing at $\mu=2J$ and well-separated two-band structure at higher detuning values. Panel (e) shows the $\mathcal{PT}$ threshold $\gamma_\mathrm{th}/J$ as a function of the gain location $m_0$ and the superconducting order parameter $\delta/J$. Near $\delta=0$, we recover the characteristic U-shaped behavior with a robust threshold $\gamma_\mathrm{th}\sim J$ when $m_0=1$, i.e. the farthest gain and loss pairs.  In contrast, for closest gain-loss locations, i.e. $m_0=(N-1)/2$, the threshold reaches $\gamma_\mathrm{th}\sim J/2$~\cite{joglekar2010,joglekar2011,agarwal2018}. This behavior is seen across the entire range of $\mu/J$; panels (f)-(h). 

In a sharp contrast, the behavior of the threshold $\gamma_\mathrm{th}/J$ as a function of the superconducting order parameter $\delta/J$ is markedly different for the zero detuning case, panel (e), vs. the nonzero detuning case, panels (f)-(h). For the latter, the threshold shows a non-monotonic suppression of $\gamma_\mathrm{th}$ with increasing $\delta/J$. When $\mu=0$, on the other hand, we see that the $\gamma_\mathrm{th}$ increases with $\delta/J$, thereby strengthening the $\mathcal{PT}$-symmetric phase. We emphasize that when $m_0=1$---gain and loss localized on the end sites---this enhancement occurs even at $\delta/J=1$. Recall that at $\delta/J=1$, the Hermitian band structure forms flat bands (Figs.~\ref{fig:PTeven}a,~\ref{fig:PTodd}a), leading to a zero threshold irrespective of $m_0$ in the even chain, Fig.~\ref{fig:PTeven}e. 
%{\bf Need a 2-3 sentence explanation of why this fails for odd $N$ and $m_0=1$.} 

%----------------------------------- Section 4 ------------------------------------------%

\section{\label{sec:N5case} Analytical approach}

\begin{figure}[h]
 \centering
 \includegraphics[width=1\columnwidth]{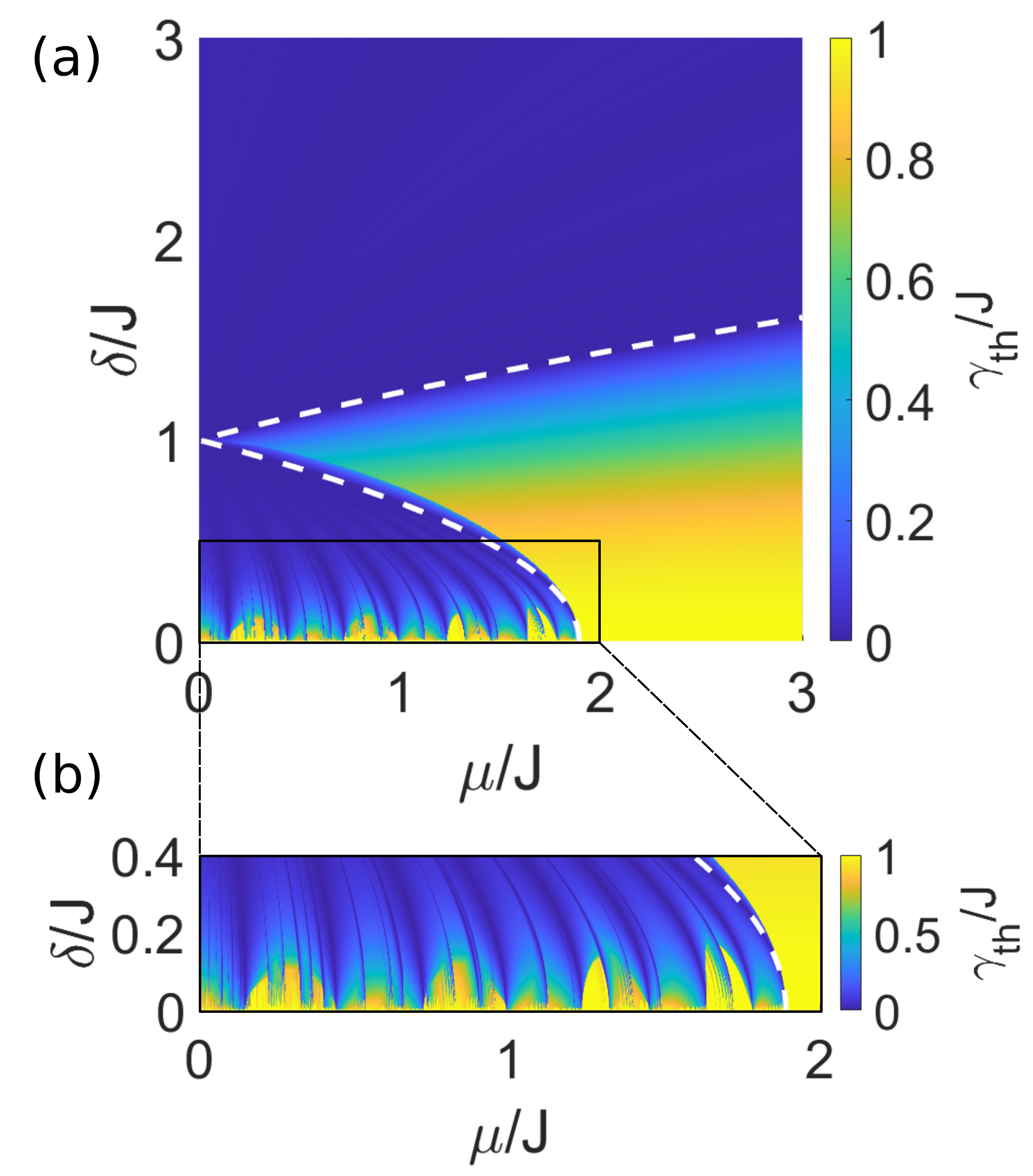}
 \caption{ (a) $\mathcal{PT}$-symmetry threshold for a chain with $N=20$ and $m_0=N/2$. The white dashed line, separating the zero-threshold region from the nonzero-threshold region, is empirically fit by equation $\alpha\mu J +|J^2 - \delta^2|=0$ where $\alpha\rightarrow 0.5$ as $N\rightarrow\infty$; at $N=20$, we find that $\al = 0.53$. This functional dependence can be obtained by requiring that two adjacent levels in the Hermitian band-structure become degenerate to get $\gamma_\mathrm{th}=0$. (b) closeup of the boxed region near the origin shows multiple ripples in $\gamma_\mathrm{th}$. }
 \label{fig:N20threshold}
\end{figure}

To get better insights into the rich structure of the $\mathcal{PT}$ threshold, we consider the behavior of $\gamma_\mathrm{th}/J$ for nearest-neighbor gain-loss potentials, $m_0=N/2$, as a function of $\mu/J$ and $\delta/J$ for an $N=20$ site chain (Fig.~\ref{fig:N20threshold}(a)).  Apart from the nonzero threshold that occurs in the limit of a non-superconducting, tight-binding chain ($\delta=0$) for any detuning, we see that $\gamma_\mathrm{th}=0$ for large $\delta$ for any $\mu$, and there is beak-shaped region in the $\mu-\delta$ plane with a positive $\mathcal{PT}$ threshold. In the magnified view of the region at small $\delta/J<1$ (Fig.~\ref{fig:N20threshold}(b)), we see significant variations in the $\PT$ threshold as we sweep across $\mu/J$. These threshold ``dips'' occur at values of $\mu/J$ where the lowest energy levels in the bulk become degenerate. The white dashed line in Fig.~\ref{fig:N20threshold}(a), separating the zero-threshold region from the positive-threshold regions, is described by equation $\al\mu J + |J^2 - \delta^2|$ where $\al\sim0.5$ is an $N$-dependent constant. The region $0<\mu/J<2$, $<\delta/J<1$ enveloped in the $\PT$ phase boundary shows many ripples with $\gamma_\mathrm{th}>0$  but it decays to zero in the thermodynamic limit. 

% Figure 4---------------------------------%
%----------------------------------------%

We remind the reader that in the $\delta=0$ tight-binding case with nearest-neighbor gain-loss potentials, all states contribute pairwise to the $\mathcal{PT}$-symmetry breaking~\cite{joglekar2011}. In contrast, in the current set up, only the states near the band edges become degenerate and then complex conjugate. In order to find the asymptotic behavior of the zero-threshold line, we turn to the Hermitian band structure, Eq.~\ref{eq:bulk}. A zero threshold is a result of degeneracy in the consecutive levels, i.e. 
$E(q_k)=E_(q_{k-1})$ where $q_k=\pi k/(N+1)$ are the lattice quasimomenta consistent with open boundary conditions. Simplifying the degeneracy criterion gives
\begin{align}
& a_1\mu J + a_2( J^2-\delta^2)=0,\\
& a_1=\cos(q_k)-\cos(q_{k-1}),\\
& a_2=a_1\left[\cos(q_k)+\cos(q_{k-1})\right].
\end{align}
Defining $\al=a_1/|a_2|$, we obtain an analytical expression for asymptotic value of $\alpha$. From the energy spectra in Figs.~\ref{fig:PTeven} (a)-(d) and numerical analysis, it follows that regions near $q\sim 0,\pi$ contribute giving $\alpha\rightarrow 0.5$ in the limit $N\rightarrow\infty$. 

%  Figure 5------------------------------%
\begin{figure*}[ht!]
    \centering
    \includegraphics[width=1\textwidth]{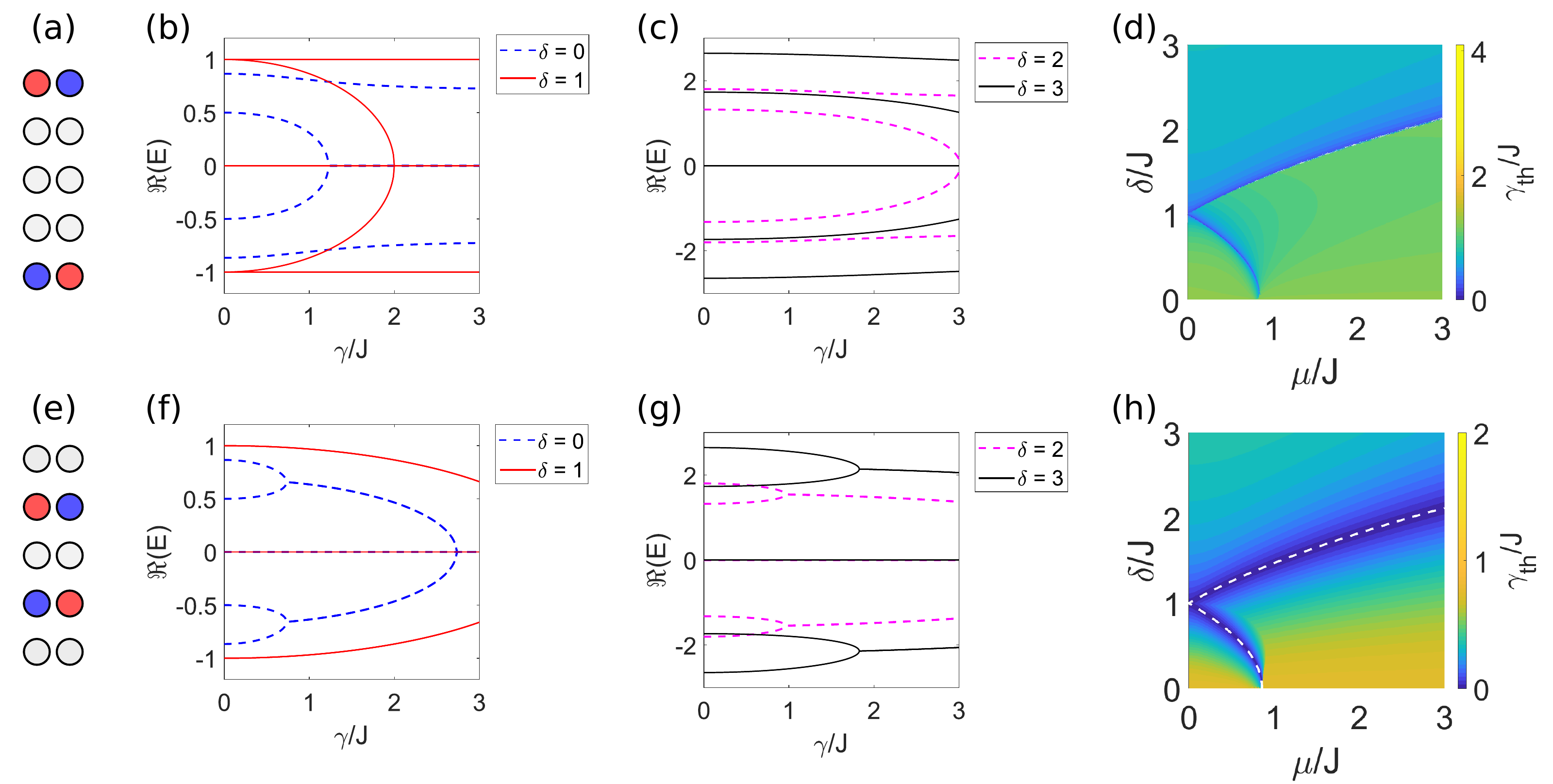}
    \caption{$\mathcal{PT}$ threshold for an $N=5$ chain. (a) Schematic of $N=5$ chain with $m_0=1$. (b)-(c) Flow of real part of eigenavalues of the $N=5$ chain as a function of $\gamma$ for farthest gain-loss locations (far left) shows that $\mathcal{PT}$ breaking occurs at an EP3 and the threshold increases monotonically with the superconducting order parameter $\delta$. (d) $\gamma_\mathrm{th}(\mu,\delta)$ shows behavior consistent with Fig.~\ref{fig:N20threshold} including a contour of zero threshold given by $\alpha\mu J+|J^2-\delta^2|=0$. (e) Schematic of $N=5$ chain with $m_0=2$. (f)-(g) Corresponding results for closest gain-loss locations, $m_0=2$, show that $\mathcal{PT}$-breaking occurs at an EP2, and the threshold varies non-monotonically with $\delta$. (h) $\gamma_\mathrm{th}(\mu,\delta)$ map shows features similar to those in panel (d).  The white-daashed line is zero-threshold contour given by $\alpha\mu J+|J^2-\delta^2|=0$ with $\alpha=1.6$.}
    \label{fig:N5EnergyThreshold}
\end{figure*}
%---------------------------------------%

Next, to understand the global behavior of the $\mathcal{PT}$ threshold $\gamma_\mathrm{th}(m_0,\mu,\delta)$ in an odd chain, we look towards the smallest nontrivial case with zero detuning, i.e. $N=5$ and $\mu=0$. When $m_0=1$, the doubly-degenerate energy spectrum is analytically tractable and is given by 
\begin{align}
 E_n &=\frac{\mp 1}{2\sqrt{2}}\left[4(J^2+\delta^2)-\gamma^2\pm\sqrt{4(J^2-\delta^2)^2+\gamma^4}\right]^{1/2},
\label{eq:m01}
\end{align}
along with two (degenerate) zero eigenvalues, $E_{5,6}=0$. As $\gamma$ is increased, the energy levels $E_{3,4}=-E_{7,8}$ first approach each other, merge with the zero-levels, and then become complex conjugate, thereby giving rise to an exceptional point of order three (EP3). The $\mathcal{PT}$-threshold in this case is given by 
\begin{equation}
\gamma_\mathrm{th}(1)=J\left[\frac{3(\delta^2+J^2)^2+4\delta^2 J^2}{2J^2(\delta^2 +J^2)}\right].
\label{eq:N5m1threshold}
\end{equation}
A similar analysis for the case with next-nearest-neighbor gain-loss potentials gives particle-hole symmetric, doubly degenerate spectra
\begin{align}
E_n&=\frac{\mp 1}{2\sqrt{2}}\left[4(J^2+\delta^2)-\gamma^2\pm\sqrt{A}\right]^{1/2},\\
A&=4(J^2-\delta^2)^2+\gamma^4-8\gamma^2(J^2+\gamma^2),
\end{align}
along with two (degenerate) zero eigenvalues, $E_{5,6}=0$. As $\gamma$ is increased, we now find that the levels near the band-edge approach each other and become degenerate, giving rise to an EP2. The $\mathcal{PT}$ threshold, obtained by requiring $E_{1,2}(\gamma_\mathrm{th})=E_{3,4}(\gamma_\mathrm{th})$, is given by 
\begin{equation}
\gamma_\mathrm{th}(2)=\left[ 4(J^2+\delta^2)-2\sqrt{ 3\delta^4+10\delta^2J^2+3J^4}\right]^{1/2}.
  \label{eq:N5m2threshold}
\end{equation}
We note that these analytical results are only valid for zero detuning, and for finite detuning $\mu>0$, we have to resort to numerical calculations. 

Figure~\ref{fig:N5EnergyThreshold}(a) shows the schematic of an $N=5$ site chain with gain-loss potentials at its ends, i.e. $m_0=1$. Panels (b)-(c) show the flow of the real parts of energy eigenvalues for the model as a function of $\gamma/J$ for different values of $\delta$. We see that increasing $\gamma$ leads to $\mathcal{PT}$-breaking that occurs at the center of the band, giving rise to an EP3. They also show that the threshold increases monotonically with $\delta$, consistent with what is seen in Fig.~\ref{fig:PTodd}e. Panel (d) shows numerically obtained threshold diagram in the $\mu-\delta$ plane.  

Figure~\ref{fig:N5EnergyThreshold}(e) shows the configuration with nearest-possible gain-loss potentials, i.e. $m_0=2$. Panels (f)-(g) show the flow of real part of eigenvalues for the model. Increasing $\gamma$ in this case leads to $\mathcal{PT}$-breaking at the band edges, and it has a non-monotonic dependence on the superconducting order parameter $\delta$, also seen in Fig.~\ref{fig:PTodd}e. Panel (h) shows numerically obtained threshold $\gamma_\mathrm{th}(\mu,\delta)$. The similarity of these threshold maps with Fig.~\ref{fig:N20threshold} is striking. By fitting the zero threshold contour to the form $\alpha\mu J+ |J^2-\delta^2|=0$ (dashed white line in panel (h)), we obtain $\alpha=1.6$. 

%----------------------------------------- Section 5 -----------------------------------%

\section{\label{sec:reentrant} Exceptional Lines and rentrant $\mathcal{PT}$ phase}
%\section{\label{sec:reentrant} Rentrant $\mathcal{PT}$ phase}

In one-dimensional lattice models with a single pair of gain and loss potentially, typically, the $\mathcal{PT}$-symmetry breaking occurs monotonically with increasing gain-loss strength $\gamma$. This is true for uniform chains with open~\cite{joglekar2010} or periodic boundary conditions~\cite{scott2012}; the Su-Schrieffer-Heeger, the Aubrey-Andre-Harper or quasi-periodic models~\cite{harter2016,harter2018}; and models with non-uniform, parity-time symmetric tunneling profiles~\cite{joglekar2011.2}, including the perfect-state transfer models. On the other hand, the presence of two or more gain-loss potentials can lead to re-entrant $\mathcal{PT}$-symmetric phase~\cite{joglekar2012,liang2014} where increasing gain-loss strength leads to repeated $\mathcal{PT}$-symmetry breaking and $\mathcal{PT}$-symmetry restoration transitions. 

% Figure 6----------------------------%
\begin{figure}[h]
    \centering
    \includegraphics[width=\columnwidth]{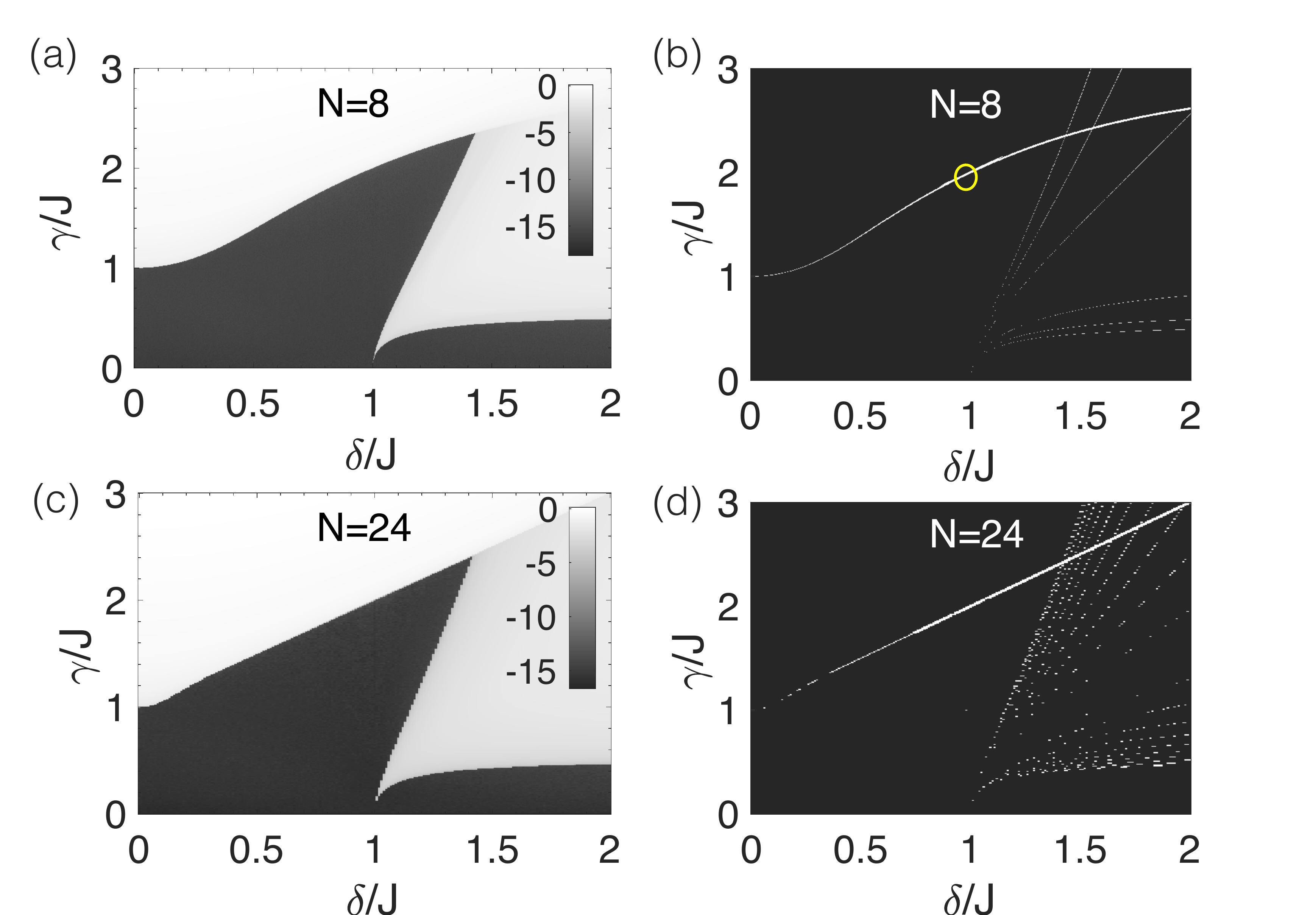}
    \caption{(a) $\mathcal{PT}$ phase diagram in the $\gamma-\delta$ plane for an $N=8$ lattice with $\mu=0$ and $m_0=1$ shows the heat map of $\Lambda\equiv\log_{10}\max_k\Im(E_k)$ where $E_k$ are the $2N$ are the eigenvalues of $H_\mathrm{K}$, Eq.(\ref{eq:hkitaev}). A re-entrant $\mathcal{PT}$-symmetric phase (black) emerges in the range $1\leq\delta/J\leq\sqrt{2}$ as the gain-loss strength $\gamma/J$ is increased. (b) EP2 contours at the $\mathcal{PT}$ boundary and in the $\mathcal{PT}$-broken region show sequential coalescence of eigenvalues. At $\delta/J=1$, due to the presence of robust Majorana modes, a third-order EP emerges at $\gamma/J=2$ (yellow circle). (c)-(d) corresponding results for an $N=24$ lattice shows same qualitative features.}
    \label{fig:PTreentrant}
\end{figure}
%------------------------------------%

% Figure 7----------------------------%
%\begin{figure*}
%    \centering
%    \includegraphics[width=\textwidth]{EPN8Eflow.pdf}
%    \caption{EP2 to EP3 change along the $\mathcal{PT}$-boundary for an $N=8$ chain. (a) Flow of $\Re(E_k)/J$ as a function of $\gamma/J$ at superconducting order parameter $\delta/J=0.7$ shows the coalescence of the midgap state and lowest-energy state in the top band. (b) EP2 contours at the $\mathcal{PT}$ boundary and in the $\mathcal{PT}$-broken region are shown in red. A locus of EP3s (blue) emerges along $\mathcal{PT}$ phase boundary for intermediate values of $\delta$. (c) Flow of $\Re(E_k)/J$ at a slightly larger value of $\delta/J=0.92$ shows that the $\mathcal{PT}$ transition is now governed by the coalescence of mid-gap state with particle-hole symmetric states, giving rise to an EP3.}
%    \label{fig:N8flow}
%\end{figure*}
%------------------------------------%

In contrast to these models with multiple non-Hermitian terms~\cite{joglekar2012,liang2014}, the Kitaev chain we have considered shows a re-entrant $\mathcal{PT}$-symmetric phase and its subsequent breaking when the gain-loss strength $\gamma$ is increased. This phenomenon occurs for an even chain with $\mu=0$ and $m_0=1$, at moderate superconducting order parameter $1\leq \delta/J\leq \sqrt{2}$, independent of the chain size. In Fig.~\ref{fig:PTreentrant}(a), we plot $\Lambda(\gamma,\delta)=\log_{10}\max_k\Im(E_k)$ where $E_k$ are (purely real or complex-conjugate) eigenvalues of the Hamiltonian $H_\mathrm{K}$ for an $N=8$ chain. The $\mathcal{PT}$-symmetric region is marked by black,  and the rest is $\mathcal{PT}$-symmetry broken region. With $\delta/J\sim 1$, as $\gamma$ is increased, the first $\PT$ symmetry breaking near $\gamma/J\sim 0.5$ occurs due to the level-attraction between and coalescence of two highest energy states in the upper band; recall that due to the particle-hole symmetric nature of the spectrum, two lowest energy levels in the lower band concurrently become degenerate. With increasing $\gamma$, subsequent lower energy levels, except the lowest state in the upper band, coalesce in pairs. This sequence of transitions leads to a large number of exceptional points in the $\mathcal{PT}$-symmetry broken region. Further increasing $\gamma$ leads to a reverse process where levels with complex-conjugate energies undergo level-attraction and $\mathcal{PT}$-symmetry is restored. For the lowest-energy states in the upper band (and their chiral counterparts), the re-entrant $\mathcal{PT}$-symmetric phase is accompanied by a qualitative change where the wave-function weight shifts from the bulk to the edges. As $\gamma$ is increased further, the system enters $\mathcal{PT}$-broken region again. This second $\mathcal{PT}$ transition across an EP is driven by coalescence of the near-zero-energy state with state at the bottom of the top band. 

To map out the exceptional point contours in the $\delta-\gamma$ plane, we use the (Dirac) inner-product matrix $M_{pq}=|\langle \psi_p|\psi_q\rangle|$ where $|\psi_k\rangle$ is the (Dirac)-normalized right eigenvector of $H_\mathrm{K}$ with eigenvalue $\lambda_k$. The order of the EP is then given by $\max_{p}\sum_{q\neq p}M_{pq}$. Figure~\ref{fig:PTreentrant}(b) shows the contours of exceptional points in the parameter space. In addition to the boundaries of $\mathcal{PT}$-symmetric and $\mathcal{PT}$-broken regions, seen in Fig.~\ref{fig:PTreentrant}(a), we see EP contours that denote the cascades of eigenvalue coalescence that occur in the $\mathcal{PT}$-broken region as $\gamma$ is increased. Of particular interest is the contour that starts at $\delta=0$ and $\gamma/J=1$. At point $\delta/J=1$, the system has fully degenerate bands with robust, mid-gap edge states (Fig.~\ref{fig:PTeven}a). Therefore, introduction of the gain-loss potentials leads to a third-order EP at $\gamma/J=2$ (shown by a yellow circle) in the otherwise second-order EP contour. We note that the prominent reentrant $\mathcal{PT}$ phases only occur when the gain-loss potentials are farthest apart, i.e. $m_0=1$ and remain robust only at $\mu=0$ for any even $N$; Figs.~\ref{fig:PTreentrant}(c)-(d) show the phase diagram and EP contours for an $N=20$ Kitaev chain.

\section{\label{sec:conclusion}Conclusion}
In this paper, we have investigated the dependence of the $\mathcal{PT}$-threshold $\gamma_\mathrm{th}$ on the properties of the underlying Hermitian Kitaev model and gain-loss potential locations. We have shown that the threshold profile is rich, with persistent differences between even and odd parity lattices. In particular, we have found that for a zero-detuning chain with odd number of sites, the threshold is enhanced with increasing superconducting order parameter. For an even chain with edge gain-loss potentials and superconducting coupling $\delta \gtrsim 1$, we discover re-entrant $\mathcal{PT}$-symmetric phase, and $\mathcal{PT}$-phase boundaries that contain both second and third order EPs. We have also discussed, briefly, a potential realization of our lattice model with coupled optical resonators. Our results further the understanding of non-Hermitian condensed matter models in the presence of realistically achievable gain and loss. 

%-------------------------------------------------------------------------------------%

%\newpage
\bibliographystyle{apsrev4-2}
\bibliography{PTKitaevBib}% Produces the bibliography via BibTeX.

%-------------------------------------------------------------------------------------%

\end{document}